\documentclass[epj]{svjour}
\usepackage{graphics}
\begin{document}
\title{A general thermodynamical description of the event horizon in the FRW universe}
\author{Fei-Quan Tu \and Yi-Xin Chen}
\institute{Zhejiang Institute of Modern Physics, Zhejiang University,
           Hangzhou, 310027, China}
\date{Received: date / Revised version: date}
%
\abstract{
The Friedmann equation in the Friedmann-Robertson-Walker(FRW) universe with any spatial curvature
is derived from the first law of thermodynamics on the event horizon.
The key idea is to redefine a Hawking temperature on the event horizon.
Furthermore, we obtain the evolution equations of the universe including the quantum correction
and explore the evolution of the universe in the $f(R)$ gravity. In addition, we also
investigate the generalized second law of thermodynamics in Einstein gravity and the $f(R)$ gravity.
This perspective also implies that the first law of thermodynamics
on the event horizon have a general description in respect of the evolution
of the FRW universe.
} 
\maketitle

\section{Introduction}

Since the discovery by Bardeen, Bekenstein, Hawking\cite{key-1,key-2,key-3}
in the 1970s, the relationship
between black hole physics and thermodynamics
have been generally accepted by physicists. Decades of research shows
that the formula of black hole entropy $S=A/4$ where
$A$ is the area of the horizon and the temperature $T=|\kappa|/(2\pi)$
where $\kappa$ is the surface gravity have a certain universality.
In 1995, Jacobson\cite{key-4} argued that Einstein equation could
be derived from the relation of thermodynamics(Clausius relation\footnote{This relation is also
called as the first law of thermodynamics (see, for example, Refs.\cite{key-5,key-6}),
so we use terms Clausius relation and the first law of thermodynamics interchangeably in this paper.})
and pointed out that Einstein equation is an equation of state. This
is an important discovery that there exists a deep connection between
Einstein gravity theory and thermodynamics. Besides, in a 4-dimensional
de Sitter space, analysis of quantum field theory shows that the temperature
of the horizon of spacetime
is $T=\kappa/(2\pi)=1/(2\pi R)$ and the total entropy is $S=\pi R^{2}$
where $R$ is the radius of the horizon\cite{key-7}. This
implies that there exists a closed relationship between the horizon
of spacetime and thermodynamics. Based on above research results, we know that
thermodynamics has a certain universality in describing the horizons
of spacetime.

For the dynamic black hole, Hayward\cite{key-8,key-9,key-10} introduced
the notion of trapping horizon in the 4-dimensional Einstein gravity
and showed that Einstein equation is equivalent to the unified first
law. Based on these facts, the authors\cite{key-11,key-12} generalized
these concepts to the FRW universe and investigated the relationship
between the unified first law and thermodynamics of the horizon
in the FRW universe. Especially, in Ref.\cite{key-11}, they considered
the FRW universe as a dynamical spherically symmetric spacetime and
defined a trapping horizon. In this way, they showed the equivalency between the
unified first law and thermodynamics of the apparent horizon in
the FRW universe.

In addition, regarding thermodynamics of the horizon, Padmanabhan\cite{key-13,key-14}
has shown that the field equations in Einstein gravity and Lanczos-Lovelock
gravity for a spherically symmetric spacetime can be expressed as the
thermodynamic identity $dE=TdS-PdV$, where the quantities $E$, $T$, $S$
and $V$ are related to the horizon and have the interpretation for
energy, temperature, entropy and volume.
So Clausius relation $\delta Q=TdS$ holds on the horizon.
On the other hand, in cosmology, there exists an event horizon
since the universe is in accelerated expansion according to astronomical observation.
Indeed, Li\cite{key-15} predicted the equation of state of the dark energy and resolved the cosmic
coincidence problem by introducing the event horizon in the model of holographic dark energy.
Besides, The event horizon of the universe is the largest comoving distance
from which light emitted now can ever reach the observer in the future
and very similar to the event horizon of the black hole whose thermodynamics
have been accepted generally. Therefore, it is natural and interesting to
investigate laws of thermodynamics related to the event horizon in the FRW universe.

For the researches about thermodynamics on the event horizon,
though Wang et al.\cite{key-16} claimed the event
horizon is unphysical from the point of view of the laws of thermodynamics,
Chakraborty\cite{key-6} concluded that the universe bounded by the event
horizon may be a Bekenstein system by redefining a Hawking temperature.
Based on the temperature defined in Ref.\cite{key-6}, we\cite{key-17}
investigated thermodynamics of the universe
bounded by the event horizon and dominated by the tachyon fluid and found
that there exists a good thermodynamic description in such universe.
However, the definition of the temperature on the event horizon
is not general in Ref.\cite{key-6,key-17}, and the thermodynamical description
is reasonable just in the flat universe and some models.
So does there exist a general thermodynamic description of the event horizon
in the FRW universe with any spatial curvature?
Indeed, we obtain the first law of thermodynamics on the event horizon
by redefining a Hawking temperature in Einstein gravity.

Now we can ask the question whether the first law of thermodynamics can hold
on the event horizon in other gravity theories such as the $f(R)$ gravity. In fact,
in the $f(R)$ gravity, Eling et al.\cite{key-18} have shown that the correct equation
of motion can not be obtained if one uses Hawking temperature, the entropy assumption $S=\alpha Af'(R)$
and the first law of thermodynamics. An entropy production term has
to be added to the first law of thermodynamics in order to obtain
the correct equation. Thus the $f(R)$ gravity is described by the
nonequilibrium thermodynamics of spacetime.
So the above question turns into the question whether the first law of thermodynamics on the event horizon
which is obtained by redefining the Hawking temperature in Einstein gravity can hold
in the $f(R)$ gravity.
In other words, can thermodynamics of the spacetime in the
$f(R)$ theory be described by the equilibrium thermodynamics?
Through the investigation, we find that the first law of thermodynamics on the event horizon is
also held in the $f(R)$ theory. Therefore, we may conclude that the
first law of thermodynamics on the event horizon has a general description
in respect of the evolution of the FRW universe.

The present paper is organized as follows. In Section 2, we show
that the first law of thermodynamics on the event horizon holds
by redefining a Hawking temperature. In Section 3, we derive the evolution equations of
the universe based on the first law of thermodynamics on the event horizon where the quantum
correction of the entropy is included. These evolution equations of the universe
can not be obtained just by Einstein equation, so the method of thermodynamical
description is more general. In Section 4, we study the evolution of the universe based on
the first law of thermodynamics on the event horizon in the $f(R)$ gravity. In Section 5,
we investigate the generalized second law of thermodynamics of the universe
bounded by the event horizon in Einstein gravity and the $f(R)$ gravity.
We end our paper with the conclusion in Section 6.
Throughout the paper, the Greek indices, $\mu,\nu,...$,
etc. run over $0,1,2,3$ and the units are chosen with $c=\hbar=k_{B}=1$
and the signature of the spacetime is taken as $(-,+,+,+)$.

\section{Redefinition of the Hawking temperature on the event horizon}

In the homogenous and isotropic universe, the metric can be expressed
as
\begin{equation}
ds^{2}=h_{ij}dx^{i}dx^{j}+R^{2}d\Omega_{2}^{2},
\end{equation}
where $i$, $j$ can take value $0$ and $1$, $R=a(t)r$ in which
$a(t)$ is the scale factor and the 2-dimensional metric $h_{ij}=diag(-1,a^{2}/(1-kr^{2}))$
in which $k$ is the spatial curvature constant. A scalar quantity
is defined as
\begin{equation}
\chi=h^{ij}\partial_{i}R\partial_{j}R.
\end{equation}

The apparent horizon is defined by the scalar quantity $\chi=0$, which
gives $R_{A}=\frac{1}{\sqrt{H^{2}+\frac{k}{a^{2}}}}$. Then the surface
gravity on the apparent horizon is defined as\cite{key-5,key-6,key-19,key-20}
\begin{equation}
\kappa_{A}=-\frac{1}{2}\frac{\partial\chi}{\partial R}\Big|_{R=R_{A}}=\frac{1}{R_{A}}
\end{equation}
and the corresponding Hawking temperature is
\begin{equation}
T_{A}=\frac{|\kappa_{A}|}{2\pi}=\frac{1}{2\pi R_{A}}.
\end{equation}
The study of thermodynamics of the apparent horizon
has made great progress in the FRW universe\cite{key-11,key-12,key-16,key-21,key-22,key-23,key-24}.
In Refs.\cite{key-11,key-12}, it was shown that the function of the
surface gravity for any horizon of the FRW universe depends on these
variables\footnote{The trapping horizon coincides with the apparent horizon $R_{A}$
in the context of the FRW universe, so we use $R_{A}$ to denote the radius of the trapping horizon.}
$R_{A}$ and $\dot{R}_{A}$  and is related
to the ratio $\dot{R}_{A}/R_{A}$ under the frame of the unified
first law.  On the other hand, Bousso\cite{key-25} pointed out that
a thermodynamic description of the horizon would be approximately
valid and it does not matter whether one uses the apparent or the event
horizon in the quintessence dominated spacetime(Q-spacetime).
Therefore, we assume that the surface gravity on the event horizon (with the
radius $R_{E}$) should have the following form
\begin{equation}
\kappa_{E}=-\frac{1}{2}\frac{\partial\chi}{\partial R}\Big|_{R=R_{E}}\frac{\dot{R}_{A}}{R_{A}}g(R_{E}),
\end{equation}
where $g(R_{E})$ is the function which is related to the variable
$R_{E}$.

Now let's determine the form of the function $g(R_{E})$. In the model
of the flat Q-spacetime(the scale factor $a(t)$ is $t^{\alpha}(\alpha>1)$
and the spatial curvature constant $k$ is $0$), the radius of the apparent horizon is $R_{A}=\frac{t}{\alpha}$
and the radius of the event horizon is $R_{E}=\frac{t}{\alpha-1}$, and
the surface gravity
on the event horizon can be reduced to the following form\cite{key-6}
\begin{equation}
\kappa_{E}=-\frac{1}{2}\frac{\partial\chi}{\partial R}\Big|_{R=R_{E}}.
\end{equation}
So the simplest form of the
function $g(R_{E})$ is
\begin{equation}
g(R_{E})=\frac{R_{E}}{\dot{R}_{E}}.
\end{equation}

Up to now, we obtain the surface gravity on the event horizon
\begin{equation}
\kappa_{E}=-\frac{1}{2}\frac{\partial\chi}{\partial R}\Big|_{R=R_{E}}\frac{\dot{R}_{A}}{R_{A}}\frac{R_{E}}{\dot{R}_{E}}.
\end{equation}
According to the relation between Hawking temperature and the surface
gravity on spacetime horizons, we get the temperature on the event
horizon
\begin{equation}
T_{E}=\frac{|\kappa_{E}|}{2\pi}=\frac{H}{2\pi}(\frac{k}{a^{2}}-\dot{H})\frac{R_{E}^{2}}{\dot{R}_{E}}.
\end{equation}

Now we would like to show the universality of this temperature on
the event horizon. The energy flux across the event horizon during
an infinitesimal time interval $dt$ can be calculated as\cite{key-6,key-16,key-24,key-25}
\begin{equation}
\delta Q=AT_{\mu\nu}k^{\mu}k^{v}dt\mid_{r=R_{E}},
\end{equation}
where $k^{\mu}$ is a null vector and $T_{\mu\nu}=(\rho+p)u_{\mu}u_{\nu}+pg_{\mu\nu}$
is the energy-momentum tensor. Thus, we can get the energy flux
\begin{equation}
\delta Q=4\pi R_{E}^{3}H(\rho+p)dt.
\end{equation}
Using the Friedmann equation $\dot{H}-\frac{k}{a^{2}}=-4\pi G(\rho+p)$,
the energy flux turns into
\begin{equation}
\delta Q=\frac{HR_{E}^{3}}{G}(\frac{k}{a^{2}}-\dot{H})dt.
\end{equation}

On the other hand, we use the Bekenstein entropy-area relation and
get
\begin{equation}
T_{E}dS_{E}=\frac{H}{2\pi}(\frac{k}{a^{2}}-\dot{H})\frac{R_{E}^{2}}{\dot{R}_{E}}\cdot2\pi R_{E}dR_{E}=\frac{HR_{E}^{3}}{G}(\frac{k}{a^{2}}-\dot{H})dt.
\end{equation}
From Eq.(12) and Eq.(13), we can see the first law of thermodynamics
$\delta Q\mid_{R_{E}}=T_{E}dS_{E}$ holds on the event horizon. In return,
we can also obtain the Friedmann equation in the FRW universe with
any spatial curvature based on the first law of thermodynamics. This is
an important result describing the event horizon of the universe.

There are some comments we would like to make regarding the thermodynamic
description on the event horizon. Firstly, obtaining Eq.(8) is based
on the following clues: (i) The FRW universe is a dynamic spherically
symmetric spacetime, so the horizon of its spacetime should be related
to the trapping horizon(analogous to the dynamic black hole).
(ii) Furthermore, the surface gravity defined
under the frame of the unified first law\cite{key-11,key-12} is related
to the ratio $\dot{R}_{A}/R_{A}$,
so it is reasonable that we assume the surface gravity on the event
horizon is related to the ratio $\dot{R}_{A}/R_{A}$ in the frame
of Clausius relation.
(iii) The concept of the event horizon of the FRW universe is
similar to that of the black hole whose thermodynamics
have been accepted generally, so the event horizon of the universe
should be described by thermodynamics.

Secondly, the form of Eq.(8) is simple and can
be reduced to the form which has been obtained in Ref.\cite{key-8}
in the model of the flat Q-spacetime, so Eq.(8) is the correct choice
and has the physical explanation of the surface gravity.

Thirdly, we get the conclusion that the first law of thermodynamics holds on
the event horizon based on Eq.(8). This is an important result which
shows the equivalency between the first law of thermodynamics on the
event horizon and Einstein equation in Einstein gravity. Because of
the conceptual similarity between the black hole horizon and the event
horizon of the universe, if we accept the thermodynamic description
of the black hole horizon, then we should agree with
the thermodynamic description of the event horizon of the universe.
What's more, we have successfully constructed the Hawking temperature
and shown the validity of the first law of thermodynamics on the event
horizon in the FRW universe.

\section{Evolution of the universe  based on the first law of thermodynamics
on the event horizon including the quantum correction}

By redefining Hawking temperature(Eq.(9)), we confirm the validity
of the first of thermodynamics on the event horizon in the above section.
In the following sections, we take the first law of thermodynamics
$\delta Q=TdS$ on the event horizon as the fundamental starting point
to derive the dynamic evolution equations of the universe.

In this section, we will consider the quantum correction of the entropy
of the event horizon and derive these evolution equations of the
universe including quantum correction effects.

As we have pointed out in Introduction, the property of the event horizon of
spacetime is similar to that of black hole. Due to the similarity,
we take the form of the quantum corrected entropy of black hole as
the entropy of the event horizon\cite{key-26,key-27,key-28,key-29,key-30}
\begin{equation}
S=\frac{A}{4L_{p}^{2}}+\alpha\ln(\frac{A}{4L_{p}^{2}}),
\end{equation}
where $\alpha$ is a constant and $L_{p}=\sqrt{\hbar G/c^{3}}$ is the Planck length. According to Ref.\cite{key-26}, $\alpha\sim O(1)$.
Thus we obtain
\begin{equation}
TdS=\frac{HR_{E}^{3}}{G}(1+\frac{\alpha L_{p}^{2}}{\pi R_{E}^{2}})(\frac{k}{a^{2}}-\dot{H})dt,
\end{equation}
and the energy flux is
\begin{equation}
\delta Q=4\pi R_{E}^{3}H(\rho+p)dt.
\end{equation}
Based on the first law of thermodynamics $\delta Q=TdS$, we get
\begin{equation}
(\frac{k}{a^{2}}-\dot{H})(1+\frac{\beta}{R_{E}^{2}})=4\pi L_{p}^{2}(\rho+p),
\end{equation}
where $\beta=\frac{\alpha L_{p}^{2}}{\pi}$ is a constant. This is the Friedmann
equation with quantum correction describing the evolution of the universe. (we will discuss it later.)

Now, in order to see the evolution properties of the universe clearly, we take the
scale factor $a(t)=t^{c}(c>1)$ and employ $G$ to denote $L_{p}^{2}$. Thus, the radius of the event horizon
turns into $R_{E}=\frac{c}{c-1}H^{-1}$ and Eq.(17) turns into
\begin{equation}
(\frac{k}{a^{2}}-\dot{H})(1+\lambda H^{2})=4\pi G(\rho+p),
\end{equation}
where $\lambda=\beta\left(\frac{c-1}{c}\right)^{2}$ is a constant.
Compared with the standard Friedmann equation, we see that this equation
has an extra term $\lambda H^{2}$ which is caused by the quantum
correction. At present, this term is very small, that's $\lambda H^{2}\ll1$,
so we can obtain
\begin{equation}
\frac{k}{a^{2}}-\dot{H}=4\pi G(\tilde{\rho}+\tilde{p}),
\end{equation}
where we redefine the effective energy density $\tilde{\rho}$ and
the effective pressure $\tilde{p}$,
\begin{equation}
\tilde{\rho}=(1-\lambda H^{2})\rho
\end{equation}
and
\begin{equation}
\tilde{p}=(1-\lambda H^{2})p,
\end{equation}
respectively. On the other hand, the continuity equation for the effective
perfect fluid is
\begin{equation}
\dot{\tilde{\rho}}+3H(\tilde{\rho}+\tilde{p})=0.
\end{equation}
Substituting  Eq.(20), Eq(21), Eq(22) into Eq.(19) and integrating
the resulting equation, we finally obtain
\begin{equation}
H^{2}+\frac{k}{a^{2}}=\frac{8\pi G}{3}(1-\lambda H^{2})\rho.
\end{equation}
This is another Friedmann equation under the quantum correction.
In order to see the properties of the accelerated expansion of the
universe clearly, we combine Eq.(19) and Eq.(23), and get the result

\begin{eqnarray}
\frac{\ddot{a}}{a} &=& -\frac{4}{3}\pi G(\rho+3p)(1-\lambda H^{2})\nonumber\\
&=& -\frac{4}{3}\pi G(\rho+3p)+\frac{4}{3}\pi G(\rho+3p)(\lambda H^{2}).
\end{eqnarray}

Comparing with the equation $\frac{\ddot{a}}{a}=-\frac{4}{3}\pi G(\rho+3p)$
which can be obtained by Einstein equation, we find Eq.(24) has an
extra term $\frac{4}{3}\pi G(\rho+3p)(\lambda H^{2})$. From the above derivation,
we know $\lambda\sim O(L_{p}^{2})$, so the extra term contains the factor $L_{p}^{2}H^{2}$
which represents quantum correction effects.

It should be noticed that the equation describing the evolution of the universe in the whole history is
Eq.(17). From this equation,
we know that the evolution of the universe depends on the event horizon $R_{E}$
and the term $\beta/{R_{E}^{2}}$ can not be ignored at the early time.
So this equation does not only show physical consistency with
classical limit but also describes quantum effects which are described by the event horizon.
Hence we can conclude that the thermodynamical description based on the event horizon under the redefinition
of Hawking temperature is more general than Einstein equation in
describing the dynamic evolution of the universe.

\section{Evolution of the universe based on the first law of thermodynamics
on the event horizon in the $f(R)$ theory }

In this section, we will investigate the evolution property of the
universe in the theory of $f(R)$ gravity. According to Eq.(10), the
energy flux is
\begin{equation}
\delta Q=4\pi R_{E}^{3}H(\bar{\rho}+\bar{p})dt,
\end{equation}
where $\bar{\rho}=\rho+\rho_{g}$ is the total energy density of the
matter energy density $\rho$ and the effective gravity energy density
$\rho_{g}$, and $\bar{p}=p+p_{g}$ is the total pressure of the matter
pressure $p$ and the effective gravity pressure $p_{g}$. In this
gravity theory the relation of entropy-area\cite{key-5,key-31} is
\begin{equation}
S=\frac{Af'(R)}{4G}.
\end{equation}
Hence
\begin{equation}
TdS=f'(R)H(\frac{k}{a^{2}}-\dot{H})\frac{R_{E}^{3}}{G}dt.
\end{equation}
Based on the first law of thermodynamics, we get the following
equation
\begin{equation}
(\frac{k}{a^{2}}-\dot{H})f'(R)=4\pi G(\bar{\rho}+\bar{p}).
\end{equation}
However, $\bar{\rho}$ and $\bar{p}$ can't be determined just by
the first law of thermodynamics. So this evolution equation of the
universe can't be also determined just by thermodynamics alone.

In order to determine the total energy density $\bar{\rho}$ and the
total pressure density $\bar{p}$, we employ the variational principle.
In the $f(R)$ theory, the Einstein-Hilbert action can be written
as
\begin{equation}
S=\int d^{4}x\sqrt{-g}\left(f(R)+2\kappa^{2}L_{m}\right),
\end{equation}
where $\kappa^{2}=8\pi G$. We employ $f$ to denote the function $f(R)$ in the following content.
Using the variational principle $\delta S=0$,
we obtain
\begin{equation}
G_{\mu\nu}=\kappa^{2}\left(\frac{1}{f'}T_{\mu\nu}^{(m)}+\frac{1}{8\pi G}T_{\mu\nu}^{(g)}\right)\equiv\kappa^{2}T_{\mu\nu}.
\end{equation}
where $G_{\mu\nu}=R_{\mu\nu}-\frac{1}{2}g_{\mu\nu}R$ is the Einstein
tensor, $T_{\mu\nu}^{(m)}=(\rho+p)u_{\mu}u_{\nu}+pg_{\mu\nu}$ is
the energy-momentum tensor of the matter, and
\begin{equation}
T_{\mu\nu}^{(g)}=\frac{1}{f'}\left[\frac{f-Rf'}{2}g_{\mu\nu}+\nabla_{\mu}\nabla_{\nu}f'-g_{\mu\nu}\nabla^{2}f'\right]
\end{equation}
is the energy-momentum tensor of the gravity. Then we get the effective
gravity energy density $\rho_{g}$ and the effective gravity pressure
$p_{g}$
\begin{equation}
\rho_{g}=\frac{1}{8\pi G}\left(\frac{Rf'-f}{2}-3Hf''\dot{R}\right)
\end{equation}
and
\begin{equation}
p_{g}=\frac{1}{8\pi G}\left(\frac{f-Rf'}{2}+f'''\dot{R}^{2}+f''\ddot{R}+2Hf''\dot{R}\right),
\end{equation}
respectively.

Thus, substituting  Eq.(32) and Eq(33) into Eq.(28), we finally get
the Friedmann equation in the FRW universe
\begin{equation}
(\frac{k}{a^{2}}-\dot{H})f'+\frac{1}{2}(Hf''\dot{R}-f'''\dot{R}^{2}-f''\ddot{R})=4\pi G(p+p).
\end{equation}
On the other hand, the continuity equation for the effective perfect
fluid in the $f(R)$ gravity is
\begin{equation}
\dot{\bar{\rho}}+3H(\bar{\rho}+\bar{p})=0.
\end{equation}
Combining Eq.(34) and Eq.(35), we obtain another Friedmann equation
\begin{equation}
H^{2}+\frac{k}{a^{2}}=\frac{8\pi G}{3f'}\left[\rho+\frac{1}{8\pi G}\left(\frac{Rf'-f}{2}-3Hf''\dot{R}\right)\right].
\end{equation}
These Friedmann equations Eq.(34) and Eq.(36) are the same as those
of Refs.\cite{key-5,key-32} which describe the evolution of
the universe in other ways. Therefore, the equivalency
between the first law of thermodynamics on the event horizon and Friedmann
equations of the FRW universe with any spatial curvature holds not
only in Einstein gravity but also in the $f(R)$ theory. This implies
that the thermodynamical description is general in describing the evolution
of the universe. Besides, it is also indicated that the FRW universe can be described
by the equilibrium thermodynamics on the event horizon in the $f(R)$ gravity.

\section{Generalized second law of thermodynamics of the universe bounded by the event horizon}

The generalized second law of thermodynamics of the universe bounded by the event horizon
in Einstein gravity has been investigated in Ref.\cite{key-33}, in which the authors
assume that the universe can be described by the equilibrium thermodynamics. But in this paper,
we have shown the validity of the first law of thermodynamics on the event horizon in Section 2, namely
the universe bounded by the event horizon can be described by the equilibrium thermodynamics.
This conclusion is particularly important for the $f(R)$ gravity, because it has been pointed
out that the spacetime in the $f(R)$ gravity is described by the nonequilibrium thermodynamics
if one uses the usual Hawking temperature\cite{key-18}.
Next, using the method of Ref.\cite{key-33}, we will present the generalized second law of
equilibrium thermodynamics of the universe bounded by the event horizon in Einstein gravity and the $f(R)$ gravity.

For the holographic dark energy (DE) model\cite{key-15} the density of holographic
DE of the universe bounded by the event horizon is
\begin{equation}
\rho_{D}=\frac{3c^{2}}{8\pi G}R_{E}^{-2},
\end{equation}
 where $c$ is a numerical constant. And the equation of state of
holographic DE can be written as
\begin{equation}
p_{D}=\omega_{D}\rho_{D},
\end{equation}
where $p_{D}$ is the thermodynamic pressure of the holographic DE
and $\omega_{D}$ is not necessarily a constant.

The two components in the matter system are non-interacting, so they satisfy
the energy conservation equations
\begin{equation}
\dot{\rho}_{d}+3H\rho_{d}=0
\end{equation}
and
\begin{equation}
\dot{\rho}_{D}+3H(\rho_{D}+p_{D})=0,
\end{equation}
separately, where $\rho_{d}$ is the energy density of dust matter (for the dust
matter, its pressure $p_{d}$ is $0$).

According to Eq.(10), the energy flux is
\begin{equation}
\delta Q=4\pi R_{E}^{3}H(\rho_{d}+\rho_{D}+p_{D})dt.
\end{equation}
In Section 2, we have shown the validity of the first law of thermodynamics
on the event horizon, so it is the equilibrium thermodynamics
and the effective temperature of the matter (dust matter and DE) distribution
can be considered to be the same as that of the event horizon\cite{key-33,key-34,key-35}.
Thus we can use the following Gibbs's relation\cite{key-34,key-35,key-36}
\begin{equation}
T_{E}dS_{m}=dE_{m}+p_{D}dV,
\end{equation}
where $S_{m}$ and $E_{m}$ are the entropy and energy of the matter
distribution. We obtain the following equation
\begin{equation}
dS_{m}=\frac{4\pi R_{E}^{2}}{T_{E}}(\rho_{d}+\rho_{D}+p_{D})dR_{E}+\frac{HR_{E}^{3}}{T_{E}}\left(\dot{H}-\frac{k}{a^{2}}\right)dt,
\end{equation}
where these relations $E_{m}=\frac{4}{3}\pi R_{E}^{3}(\rho_{d}+\rho_{D})$
and $V=\frac{4}{3}\pi R_{E}^{3}$ are used. Substituting Eq.(37) and
Eq.(38) into Eq.(40), we get
\begin{equation}
dR_{E}=\frac{3}{2}R_{E}H(1+\omega_{D})dt.
\end{equation}
Hence the change of the total entropy $S_{tot}=S_{m}+S_{E}$ where $S_{E}$ is
the entropy of the event horizon which is determined by Eq.(13) is
\begin{equation}
\frac{dS_{tot}}{dt}=\frac{6\pi R_{E}^{3}H}{T_{E}}(\rho_{d}+\rho_{D}+p_{D})(1+\omega_{D}).
\end{equation}
We see that the result is the same as that of Ref.\cite{key-33}.
When the holographic DE satisfies the weak energy condition
\begin{equation}
\rho_{D}+p_{D}=(1+\omega_{D})\rho_{D}\geq 0,
\end{equation}
the generalized second law of thermodynamics will be valid for the universe bounded by the event horizon.

For the $f(R)$ gravity, as we have shown in Section 4, the first
law of thermodynamics holds on the event horizon, so Gibbs's relation
(42) can be used. Thus we obtain
\begin{equation}
dS_{m}=\frac{4\pi R_{E}^{2}}{T_{E}}(\rho_{d}+\rho_{g}+p_{g})dR_{E}+\frac{HR_{E}^{3}}{T_{E}}\left(\dot{H}-\frac{k}{a^{2}}\right)dt,
\end{equation}
where we employ the dust as the matter. According to the definition
of the event horizon $R_{E}=a(t)\int_{t}^{\infty}\frac{dt'}{a(t')}$,
we get
\begin{equation}
dR_{E}=(HR_{E}-1)dt,
\end{equation}
so the change of the total entropy is
\begin{eqnarray}
\frac{dS_{tot}}{dt}&=&\frac{4\pi R_{E}^{2}}{T_{E}}(\rho_{d}+\rho_{g}+p_{g})(HR_{E}-1)\nonumber\\
&+&(1-f')\frac{HR_{E}^{3}}{GT_{E}}\left(\dot{H}-\frac{k}{a^{2}}\right).
\end{eqnarray}
Substituting Eq.(32) and Eq.(33) into Eq.(49), we get
\begin{eqnarray}
\frac{dS_{tot}}{dt}&=&\frac{R_{E}^{2}}{2GT_{E}}(\rho_{d}+f'''\dot{R}^{2}+f''\ddot{R}-Hf''\dot{R})(HR_{E}-1)\nonumber\\
&+&(1-f')\frac{HR_{E}^{3}}{GT_{E}}\left(\dot{H}-\frac{k}{a^{2}}\right).
\end{eqnarray}
So the generalized second law of thermodynamics can be satisfied
as long as the above expression is not less than 0.

Now we would like to make some remarks regarding the generalized second
law of thermodynamics. (i) From the above derivation, we know that
the Gibbs's relation (42) is important in order to obtain the change
of the total entropy. Indeed, we have established the Gibbs's relation on the event
horizon in Section 2, i.e. the first law of thermodynamics on the event horizon.
By contrast, the authors in Ref.\cite{key-33} just assumed the validity of the first law of thermodynamics on
the event horizon and a temperature of the event horizon whose expression is unknown.
(ii) For the $f(R)$ gravity, if one does not redefine the Hawking
temperature, then the horizon is described by the nonequilibrium thermodynamics\cite{key-11,key-18}.
As we have known, the Gibbs's relation (42) can not be used for the
nonequilibrium thermodynamics, so the method in Ref.\cite{key-33}
is invalid. However, the first law of thermodynamics on the event
horizon holds and the Gibbs's relation (42) can be used in this paper.
(iii) For the $f(R)$ gravity, the form of the change of the total entropy
is analytical, so it is convenient to discuss the generalized second
law of thermodynamics if some physical quantities are given.

\section{Conclusion}

So far, the study of thermodynamics of the event horizon is
rare while the research of thermodynamics of the apparent horizon
has made great progress in the FRW universe. However, there exists
an event horizon since the universe is in accelerated expansion. What is more,
the concept of the event horizon of the universe is very similar to the horizon of the black hole
whose thermodynamics have been accepted generally. Hence it is natural and important to study thermodynamics
of the event horizon in the FRW universe. As far as we know, the difficulty
of studying thermodynamics on the event horizon is the definition
of the temperature. For example, in Ref.\cite{key-16} the authors employed
the temperature on the event horizon $T_{E}=1/(2\pi R_{E})$ whose
form is similar to that of the apparent horizon and showed that the first
law of thermodynamics on the event horizon is out of work. In Ref.\cite{key-6},
the author redefined the Hawking temperature but the Hawking temperature
is not general, and his conclusions are only suitable for the flat
spacetime and some models.

In order to solve these difficulties, we redefine the surface gravity
and the corresponding Hawking temperature on the event horizon. Subsequently, we
show the equivalency between the first law of thermodynamics on the
event horizon and Friedmann equations of FRW universe with any spatial
curvature in Einstein gravity. That is to say, the first law of thermodynamics
on the event horizon holds in the FRW universe with any spatial
curvature in Einstein gravity. This is a very important property which
indicates the event horizon can be described by equilibrium thermodynamics.

Then, starting with the first law of thermodynamics on the event horizon,
we obtain Friedmann equation including the quantum correction
and show that the evolution of
the universe is related to the event horizon. As an example, we present the evolution
of the universe at present and get corresponding quantum
corrected Friedmann equations which are consistent with the standard
Friedmann equations under classical limit. Furthermore, we obtain
Friedmann equations of the FRW universe with any spatial curvature in
the $f(R)$ gravity based on the first law of thermodynamics.
Subsequently, we explore the generalized second law
of thermodynamics of the universe bounded by the event horizon and
get these conditions which satisfy the generalized second law
of thermodynamics in Einstein gravity and the $f(R)$ gravity.
In summary, we conclude that the first law of thermodynamics on the event
horizon has a general description in respect of the evolution of the FRW
universe.

\section*{Acknowledgements}
This work is supported  by the NNSF of China, Grant No. 11375150.
%
%

\end{document}